\documentclass[12pt]{article}
\pdfoutput =1
\usepackage{float} 
\usepackage{amsmath}
\usepackage{slashed}
\usepackage{amsfonts}   
\usepackage{amssymb}

\begin{document}
\date{}
\title{{\bf{\Large Probing tachyon kinks in Newton-Cartan background}}}
\author{
 {\bf {\normalsize Dibakar Roychowdhury}$
$\thanks{E-mail:  dibakarphys@gmail.com, dibakarfph@iitr.ac.in}}\\
 {\normalsize  Department of Physics, Indian Institute of Technology Roorkee,}\\
  {\normalsize Roorkee 247667, Uttarakhand, India}
\\[0.3cm]
}

\maketitle
\begin{abstract}
In this paper, we perform a systematic analysis of tachyon condensation over string Newton-Cartan (NC) geometry by probing the background with non BPS $ Dp $ brane. We construct the \textit{finite} tachyon effective action for NC geometry and explore the dynamics associated with the tachyon kink on the world-volume of the non BPS $ Dp $ brane. We show that under certain specific assumptions, the spatial dependent tachyon condensation leads to an emerging BPS $ D(p-1) $ brane dynamics over NC background. We further compute the strees energy tensor and show the equivalence between two different $ Dp $ brane configurations in the NC limit.
\end{abstract}
\section{Overview and Motivation}
Over the last couple of decades, the non relativistic string theory \cite{Gomis:2000bd} with world-sheet Galilean invariance has attained renewed attention due to its several remarkable features. One of the major aspects associated with the non relativistic formulation of string theory turns out to be the existence of an UV finite unitary description that governs the stringy dynamics over a Galilean invariant geometry \cite{Gomis:2004pw}-\cite{Harmark:2018cdl}. Interestingly enough, the non relativistic string spectrum does not seem to contain any massless excitation which therefore sort of guarantees that the corresponding low energy dynamics associated with that of the target space geometry is not governed by Einstein gravity. On contrary, the corresponding target space geometry turns out to be string Newton-Cartan (NC) background \cite{Andringa:2012uz}-\cite{Bergshoeff:2015uaa} which differs significantly from that of the usual Riemannian spacetime.

Recently, there has been a lot of attention towards a profound understanding of stringy dynamics over NC background \cite{Bergshoeff:2018vfn}-\cite{Harmark:2017rpg}. For example, a systematic theoretical formulation of four dimensional extended string Newton-Cartan gravity has been provided by authors in \cite{Bergshoeff:2018vfn}. However, in spite of these attempts, several crucial issues associated to string NC geometry are yet to be answered. One fundamental question that remains to be addressed is the understanding of the dynamics of tachyon kinks \cite{Sen:2003tm}-\cite{Calo:2009wu} on the world-volume of non BPS $ Dp $ branes \cite{Sen:1999md}-\cite{Garousi:2000tr} embedded in NC geometry. The purpose of the present article is therefore twofold- (i) to construct an \textit{effective} field theory for tachyons in the NC limit \cite{Bergshoeff:2015uaa} and (ii) to check whether the kink solution associated with non BPS $ Dp $ branes can give rise to an emerging $ D(p-1) $ brane dynamics over string NC geometry.

Our analysis reveals that tachyon condensation over string NC geometry differs significantly from that of the earlier analysis in several aspects. First of all, in order to have a \textit{finite} world-volume theory for non BPS $ Dp $ branes one needs to consider the effective WZ coupling to be different from that of the usual coupling associated with the DBI action. More specifically, we show that in order for the action to be finite this WZ coupling has to go \textit{linearly} with the NC parameter ($ \omega $) in the limit $ \omega \rightarrow \infty $. Secondly, unlike the previous analysis \cite{Sen:2003tm}, one needs to consider the most \textit{generic} ansatz for the scalar modes living on the world-volume of non BPS $ Dp $ brane. We show that a systematic analysis of the kink solution (subjected to a set of \textit{constraints}) performed under these two assumptions naturally leads towards an emerging $ D(p-1) $ brane dynamics in the NC ($ \omega \rightarrow \infty $) limit \cite{Bergshoeff:2015uaa}. In order to solidify our claim, we further compute the stress energy tensor(s) associated with the non BPS $ Dp $ brane configuration and show that it is exactly equal to the stress energy tensor(s) associated with the BPS $ D(p-1) $ brane dynamics once the tachyon equation of motion together with the above set of constraints are taken care of. 

The rest of the paper is organized as follows. We start with a formal construction of effective field theory for tachyons in Section 2. In Section 3, we explore the dynamics associated with tachyon kinks (over string NC geometry) and show that this dynamics is equivalent to that of the BPS $ D(p-1) $ dynamics once a certain set of constraints together with the tachyon equation of motion are imposed. To complete this analogy, we further compute stress tensor(s) in Section 4 and show that these two brane configurations are \textit{equivalent}. Finally, we conclude in Section 5.
\section{Tachyon effective action}
We start our analysis with a formal introduction to the non-BPS $Dp$ branes in NC background. The world-volume action corresponding to the bosonic sector is proposed to be of the following form\footnote{We restrict ourselves to the configuration with vanishing NS-NS flux and constant dilaton.} \cite{Sen:1999md},
\begin{eqnarray}
\mathcal{S}_{N Dp}&=&\mathcal{S}_{DBI}+\mathcal{S}_{WZ}\nonumber\\
&=&-\tau_p \int d^{p+1}\xi ~V(T)\sqrt{-| \mathcal{A}|}+ \gamma_{eff} (\omega) \int V(T)~dT \wedge C^{(p)}\wedge e^{F}\label{e1}
\end{eqnarray}
where, the individual entities could be formally expressed as,
\begin{eqnarray}
\mathcal{A}_{\alpha \beta}=\mathcal{G}_{\mu \nu}\partial_{\alpha}X^{\mu}\partial_{\beta}X^{\nu}+\mathcal{F}_{\alpha \beta}+\partial_{\alpha}T \partial_{\beta}T
\end{eqnarray}
together with the fact that, $ X^{\mu}(\mu =0,..,d-1) $ are the embedding coordinates and $ \alpha =0,...,p $ are the world volume directions. Here, $ V(T) $ is the tachyon potential with the assumption that it is even function of $ T $ that goes to zero as $ T \rightarrow \pm \infty $ and becomes maximum at $ T=0 $. Here, $ \omega $ is a free parameter that goes to infinity in the NC limit \cite{Andringa:2012uz}. We show that in order for the world-volume theory (\ref{e1}) to make sense the \textit{effective} coupling for the RR $ p $ form (in the NC limit) must go linearly with $ \omega $ namely\footnote{In principle, there could be more than one proposal for the (WZ) effective coupling associated with non BPS $ Dp $ branes in NC gravity. For example, one possible choice could have been of the form $ \gamma_{eff}(\omega) \sim \omega^{2}\sin (\omega^{-1})$ which goes like $ \omega-\frac{1}{6 \omega} +\mathcal{O}(\omega^{-2})$ in the limit $ | \omega | \gg 1 $.},
\begin{eqnarray}
\lim_{\omega \rightarrow \infty}\gamma_{eff} (\omega)\sim \tau_p ~ \omega
\end{eqnarray}
where, $ \tau_p $ is the usual DBI coupling of the theory.

Following \cite{Andringa:2012uz}-\cite{Bergshoeff:2015uaa}, we express the vierbein field as,
\begin{eqnarray}
E_{\mu}^{a}=\omega \tau_{\mu}^{a}+\frac{1}{2\omega}m_{\mu}^{a},~~E_{\mu}^{a'}=e_{\mu}^{a'}\label{E3}
\end{eqnarray}
where, $ a,b=0,..,p $ are the longitudinal (world-volume) directions (together with, $ \eta_{ab}=diag (-1,1,...,1) $) and $ a'=p+1,..,d-1 $ corresponds to transverse directions. 

With this set up in hand, it is straightforward to compute,
\begin{eqnarray}
\mathcal{G}_{\mu \nu}\partial_{\alpha}X^{\mu}\partial_{\beta}X^{\nu}=\omega^{2}\mathfrak{a}_{\alpha \beta}+\mathfrak{h}_{\alpha \beta}+\mathfrak{b}_{\alpha \beta}+\frac{1}{4\omega^{2}}\mathfrak{c}_{\alpha \beta}
\end{eqnarray}
where, individual entities could be formally expressed as,
\begin{eqnarray}
\mathfrak{a}_{\alpha \beta}&=&\tau_{\mu \nu}\partial_{\alpha}X^{\mu}\partial_{\beta}X^{\nu};~~\tau_{\mu \nu}=\tau_{\mu}^{a}\tau_{\nu}^{b}\eta_{ab}\nonumber\\
\mathfrak{h}_{\alpha \beta}&=&\mathfrak{h}_{\mu \nu}\partial_{\alpha}X^{\mu}\partial_{\beta}X^{\nu};~~\mathfrak{h}_{\mu \nu}= e_{\mu}^{a'}e_{\nu}^{b'}\delta_{a' b'}\nonumber\\
\mathfrak{b}_{\alpha \beta}&=&\mathfrak{b}_{\mu \nu}\partial_{\alpha}X^{\mu}\partial_{\beta}X^{\nu};~~\mathfrak{b}_{\mu \nu}= \frac{1}{2}\eta_{ab}\left(\tau_{\mu}^{a}m_{\nu}^{b}+m_{\mu}^{a}\tau_{\nu}^{b} \right)\nonumber\\
\mathfrak{c}_{\alpha \beta}&=&\mathfrak{c}_{\mu \nu}\partial_{\alpha}X^{\mu}\partial_{\beta}X^{\nu};~~\mathfrak{c}_{\mu \nu}=m_{\mu}^{a}m_{\nu}^{b}\eta_{ab}.\label{E5} 
\end{eqnarray}

In order to take a consistent NC limit one needs to constrain the curvature ($ \mathcal{F}_{\alpha \beta} $) corresponding to the abelian connection in a specific manner which is to set, $ \mathcal{F}_{\alpha \beta}=0$. In other words, we treat the corresponding $ U(1) $ connection as pure gauge \cite{Bergshoeff:2015uaa}. Finally, we define tachyon as well as the $ Dp $ brane tension in the NC limit as,
\begin{eqnarray}
T \rightarrow T;~~\tau_{p}\rightarrow  \tilde{\tau_{p}}= \omega^{p-1}\tau_{p}\label{e6}
\end{eqnarray} 
which yields the DBI action of the following form,
\begin{eqnarray}
\mathcal{S}_{DBI}=-\tilde{\tau_{p}}\omega^{2}\int d^{p +1}\xi~ V(T)\sqrt{- | \mathfrak{a}|}-\frac{\tilde{\tau_p}}{2}\int d^{p+1}\xi~V(T)\sqrt{- | \mathfrak{a}|}~\mathfrak{a}^{\alpha \beta}\mathfrak{g}_{\alpha \beta}+\mathcal{O}(\omega^{-2})\label{e7}
\end{eqnarray}
where, $ \mathfrak{a} $ is a $ (p+1)\times (p+1) $ non singular matrix together with,
\begin{eqnarray}
\mathfrak{g}_{\alpha \beta}(X^{\mu},\tau_{\mu}^{a},e_{\mu}^{a'},m^{a}_{\mu},T)&=&\mathfrak{h}_{\alpha \beta}(X^{\mu}, \tau_{\mu}^{a},e_{\mu}^{a'},m^{a}_{\mu})+\mathfrak{b}_{\alpha \beta}(X^{\mu}, \tau_{\mu}^{a},e_{\mu}^{a'},m^{a}_{\mu})+\partial_{\alpha}T \partial_{\beta}T\nonumber\\
& \equiv & \tilde{\mathfrak{h}}_{\alpha \beta}(X^{\mu},\tau_{\mu}^{a},e_{\mu}^{a'},m^{a}_{\mu})+\partial_{\alpha}T \partial_{\beta}T
\end{eqnarray}
where we have denoted, $ \tilde{\mathfrak{h}}_{\alpha \beta}:=\mathfrak{h}_{\alpha \beta}+\mathfrak{b}_{\alpha \beta} $. Clearly, the first term on the R.H.S. of (\ref{e7}) is divergent in the limit, $ \omega \rightarrow \infty $ which thereby needs to be fixed in order for the theory to make sense in the NC framework. On the other hand, the RR $ p $ form coupled with the non BPS $ Dp $ brane could be formally expressed as,
\begin{eqnarray}
C^{(p)}=\frac{1}{p!}\mathcal{C}_{\mu_{0}..\mu_{p-1}}dX^{\mu_0}\wedge ...\wedge dX^{\mu_{p-1}}
\end{eqnarray}
where we presume that in the NC formulation of non BPS $ Dp $ brane \cite{Bergshoeff:2015uaa},
\begin{eqnarray}
\mathcal{C}_{\mu_{0}..\mu_{p-1}}=\left(\omega\tau_{\mu_{0}}^{a_0}-\frac{1}{2\omega}m_{\mu_{0}} ^{a_0}\right)...\left(\omega\tau_{\mu_{p-1}}^{a_{p-1}}-\frac{1}{2\omega}m_{\mu_{p-1}}^{a_{p-1}}\right) \epsilon_{a_{0}...a_{p-1}}.\label{e10}
\end{eqnarray}

Using (\ref{e6}) and (\ref{e10}), it is indeed quite straightforward to show that in the limit, $ \omega \rightarrow \infty $
\begin{eqnarray}
\mathcal{S}_{WZ}=\tilde{\tau}_{p}\omega^{2}(-1)^{p}\int d^{p+1}\xi~V(T)~ \partial_{\alpha_{p}}T~ |\tau_{\alpha}^{a}|
\end{eqnarray}
where, we have used the notation, $ \tau_{\alpha}^{a}= \tau_{\mu}^{a}\partial_{\alpha}X^{\mu}$ together with the fact,
\begin{eqnarray}
|\tau_{\alpha}^{a}|=\frac{1}{p!}\epsilon^{\alpha_{0}..\alpha_{p-1}}\tau_{\alpha_{0}}^{a_{0}}..\tau_{\alpha_{p-1}}^{a_{p-1}}\epsilon_{a_0 ..a_{p-1}}.
\end{eqnarray}
Here, $ [\tau_{\alpha}^{a}] $ corresponds to a $ p \times p $ non-singular matrix. A careful look reveals that $ [\tau_{\alpha}^{a}] $ is actually a \textit{sub-matrix} ($ \tilde{\mathfrak{a}} $) within the full $ (p+1) \times (p+1) $ dimensional matrix $ \mathfrak{a} $.

Therefore, in order for the action (\ref{e1}) to be finite in the NC limit, one must impose the following constraint,
\begin{eqnarray}
\sqrt{-|\mathfrak{a}|}=(-1)^{p}\partial_{\alpha_{p}}T~ \sqrt{-| \tilde{\mathfrak{a}}|}~;~~\tilde{\mathfrak{a}}_{\alpha_m \beta_n}=\tau_{\mu \nu}\partial_{\alpha_{m}}X^{\mu}\partial_{\beta_{n}}X^{\nu};~m,n=0,..,p-1\label{E13}
\end{eqnarray}
which thereby leads to the \textit{finite} world-volume action for the non BPS $ Dp $ brane in the NC background,
\begin{eqnarray}
\mathcal{S}_{N Dp}=-\frac{\tilde{\tau_p}}{2}(-1)^{p}\int d^{p+1}\xi~V(T)\partial_{\alpha_{p}}T~ \sqrt{-| \tilde{\mathfrak{a}}|}~\mathfrak{a}^{\lambda \sigma}\mathfrak{g}_{\lambda \sigma}\equiv -\frac{\tilde{\tau_p}}{2}(-1)^{p}\int d^{p+1}\xi~ \mathcal{L}_{N Dp}.\label{e13}
\end{eqnarray}
One should notice that in the NC formulation, the WZ term plays crucial role in order to arrive at a finite effective action for tachyons. Interestingly enough, for non BPS $ Dp $ branes one pays a special price (in the NC limit) namely, the RR $ p $ form ($ C^{(p)} $) couples differently (\ref{e10}) compared to that of the usual scenario with relativistic $ p $ branes \cite{Sen:1999md}. In other words, in the NC formulation of non- BPS $ Dp $ branes one is tempted to consider a different coupling strength corresponding to the WZ term. This stems from the fact that unlike the case for BPS $ Dp $ branes, the non BPS $ Dp $ branes are always coupled with background RR $ p $ form rather than a $ (p+1) $ form \cite{Gomis:2005bj}. 

\section{Tachyon dynamics}
\subsection{Tachyon kink}
The purpose of this Section is to explore the tachyon kink solution \cite{Sen:2003tm} on the world volume of non BPS $ Dp $ brane defined above in (\ref{e13}). The Lagrangian (\ref{e13}) could be explicitly written as,
\begin{eqnarray}
\mathcal{L}_{N Dp}=V(T)\partial_{\alpha_{p}}T~\sqrt{-| \tilde{\mathfrak{a}}(\tau_{\mu \nu},X^{\beta})|}~ \mathfrak{a}^{\lambda_{j} \sigma_{k}}(\tau_{\mu \nu},X^{\beta})~(\tilde{\mathfrak{h}}_{\lambda_{j} \sigma_{k}}(\tilde{\mathfrak{h}}_{\mu \nu},X^{\beta})+\partial_{\lambda_{j}}T \partial_{\sigma_{k}}T)\label{e14}
\end{eqnarray}
where, $ j,k=0,..,p $.

The equations of motion that readily follow from (\ref{e14}) could be formally expressed as,
\begin{eqnarray}
V \frac{\partial}{\partial \xi^{p}}\left(\sqrt{-| \tilde{\mathfrak{a}}|}\mathfrak{a}^{\lambda_{j}\sigma_{k}}(\tilde{\mathfrak{h}}_{\lambda_{j} \sigma_{k}}+\partial_{\lambda_{j}}T \partial_{\sigma_{k}}T) \right)
+2\frac{\partial}{\partial \xi^{\lambda_{j}}}\left(V \partial_{\alpha_{p}} T \sqrt{-| \tilde{\mathfrak{a}}|}\mathfrak{a}^{\lambda_{j}\sigma_{k}} \partial_{\sigma_{k}}T \right)=0,
\label{e16} 
\end{eqnarray}
\begin{eqnarray}
\frac{\sqrt{-| \tilde{\mathfrak{a}}|} }{2} \left( \frac{\partial \tilde{\mathfrak{a}}_{\alpha_{m}\beta_n}}{\partial X^{\rho}}\right) \tilde{\mathfrak{a}}^{\alpha_{m}\beta_n}\mathfrak{a}^{\lambda_{j}\sigma_{k}}(\tilde{\mathfrak{h}}_{\lambda_{j} \sigma_{k}}+\partial_{\lambda_{j}}T \partial_{\sigma_{k}}T)+\sqrt{-|\tilde{\mathfrak{a}}|}\frac{\partial}{\partial X^{\rho}}  ( \mathfrak{a}^{\lambda_j \sigma_k}\tilde{\mathfrak{h}}_{\lambda_j \sigma_k})&=&0.
\label{e17}
\end{eqnarray}

In order to proceed further we choose the following ansatz \cite{Sen:2003tm} for the massless modes associated with the world-volume theory,
\begin{eqnarray}
T (\xi^{p},\xi^{m})=\mathfrak{f}(\zeta (\xi^{p}-\mathfrak{t}(\xi^{m})));~~X^{\mu}=X^{\mu}(\xi^{j});~~m=0,..,p-1
\label{e18}
\end{eqnarray}
subjected to the fact,
\begin{eqnarray}
\mathfrak{f}(-u)=-\mathfrak{f}(u);~~\mathfrak{f}'(u)>0~~\forall u;~~\mathfrak{f}(\pm \infty)=\pm \infty
\end{eqnarray}
where we set, $ \zeta =\infty $ towards the end of our calculation. In this limit, $ T= \infty $ for $ x> \mathfrak{t}(\xi^{m}) $ and $ T= -\infty $ for $ x< \mathfrak{t}(\xi^{m}) $. Here, $ \mathfrak{t}(\xi^{m}) $ is the field that propagates on the world-volume of the kink defined by means of world-volume coordinates $ \lbrace \xi^{m}\rbrace $.

Next, our task is to substitute the above ansatz (\ref{e18}) into (\ref{e16})-(\ref{e17}) and to check whether (\ref{e16})-(\ref{e17}) boil down to the set of equations that could be obtained from a world-volume theory of BPS $ D(p-1) $ brane embedded in the NC background. In order to do so, it is customary first to note down the $ (p+1)\times (p+1) $ matrix $ \mathfrak{a}_{\lambda_{j}\sigma_{k}} $ and its inverse. The components of $ \mathfrak{a}_{\lambda_{j}\sigma_{k}}  $ could be formally expressed as,
\begin{eqnarray}
\mathfrak{a}_{\lambda_{p}\sigma_{p}}& \equiv & \mathfrak{a}_{pp}=\tau^{a}_{p}\tau^{b}_{p}\eta_{ab};~~\mathfrak{a}_{p \sigma_{n}}=\tau^{a}_{p}\tau^{b}_{n}\eta_{ab};~~m,n=0,..,p-1\nonumber\\
\mathfrak{a}_{\lambda_{m}p}& = &\tau^{a}_{m}\tau^{b}_{p}\eta_{ab};~~\mathfrak{a}_{\lambda_{m}\sigma_{n}}\equiv \tilde{\mathfrak{a}}_{\lambda_{m}\sigma_{n}} =\tau^{a}_{m}\tau^{b}_{n}\eta_{ab}
\end{eqnarray}
which finally yields the inverse matrix,
\begin{eqnarray}
 \mathfrak{a}^{pp}& =& \frac{1}{\zeta^{2} \mathfrak{f}'^{2}}~;~~\mathfrak{a}^{\lambda_{m}p}=- \frac{1}{\zeta^{2} \mathfrak{f}'^{2}}\tilde{\mathfrak{a}}^{\lambda_{m}\sigma_{n}}\mathfrak{a}_{p \sigma_{n}}\nonumber\\
 \mathfrak{a}^{\lambda_{m}\sigma_{n}}& =& \frac{1}{\zeta^{2} \mathfrak{f}'^{2}}\tilde{\mathfrak{a}}^{\lambda_{m}\sigma_{n}}\mathfrak{a}_{pp}~;~~\mathfrak{a}^{p\sigma_{n}}=- \frac{1}{\zeta^{2} \mathfrak{f}'^{2}}\tilde{\mathfrak{a}}^{\lambda_{m}\sigma_{n}}\mathfrak{a}_{\lambda_{m}p}\label{e21}
\end{eqnarray}
where we have used the standard identity, $ \frac{\delta | \tilde{\mathfrak{a}}|^{-1}}{ | \tilde{\mathfrak{a}}|^{-1}}:=\tilde{\mathfrak{a}}^{-1}\circ\delta\tilde{\mathfrak{a}}$ along with the following identifications namely, $ | \tilde{\mathfrak{a}}|^{-1}=det~ \tilde{\mathfrak{a}}^{\alpha_{m}\beta_{n}}=\frac{1}{det~ \tilde{\mathfrak{a}}_{\alpha_{m}\beta_{n}}} $ and $ \tilde{\mathfrak{a}}=\tilde{\mathfrak{a}}^{\alpha_{m}\beta_{n}} $.

Using (\ref{e21}), it is now straightforward to evaluate each of the individual entities (in the limit, $ \zeta \rightarrow \infty $) in (\ref{e16}) and (\ref{e17}). We first note down the L.H.S. of (\ref{e16}). A careful analysis reveals that in the limit, $ \zeta \rightarrow \infty $ the tachyon dynamics (\ref{e16}) essentially boils down to,
\begin{eqnarray}
V'(T)\mathfrak{f}'  \sqrt{-| \tilde{\mathfrak{a}}|}\mathfrak{a}^{\lambda_j \sigma_k}\partial_{\lambda_{j}}T \partial_{\sigma_{k}}T=0.\label{ee22}
\end{eqnarray}

Given (\ref{ee22}), we now explore various consequences associated to it. Let us first consider the situation where $ | \xi^{p}- \mathfrak{t}(\xi^{m})|  $ is finite such that, $ \zeta | \xi^{p}- \mathfrak{t}(\xi^{m})|\gg 1  $. In this limit the tachyon potential ($ V(T) $) goes to its minimum which thereby implies that $ V'(\pm\infty)=0 $. On the other hand, the potential reaches its maximum at $\xi^{p}= \mathfrak{t}(\xi^{m})  $ which also sets $ V'(0)=0 $ . Therefore, we conclude that (\ref{ee22}) is identically satisfied in these two limits. However, apart from these two cases one might also consider a third example where the product $ \zeta | \xi^{p}- \mathfrak{t}(\xi^{m})|  $ remains \textit{finite} in the limit when $ \zeta \gg 1 $ and $ | \xi^{p}- \mathfrak{t}(\xi^{m})|\ll 1 $. This therefore strongly forces us to set,
\begin{eqnarray}
\mathfrak{a}^{\lambda_j \sigma_k}\partial_{\lambda_{j}}T \partial_{\sigma_{k}}T=0\label{ee23}
\end{eqnarray}
which we claim to be true for the entire range $ -\infty <T <\infty $ as the L.H.S. of (\ref{ee23}) does not depend on the parameter $ \zeta $.

Finally, we consider equation of motion (\ref{e17}) corresponding to massless scalar fluctuations ($ X^{\mu} $). A straightforward computation reveals, 
\begin{eqnarray}
\frac{1}{2}\sqrt{-|\tilde{\mathfrak{a}}|}\left(\frac{\partial \tilde{\mathfrak{a}}_{\alpha_m \beta_n}}{\partial X^{\rho}} \right) \tilde{\mathfrak{a}}^{\alpha_m \beta_n}
(\tilde{\mathfrak{a}}^{\lambda_m \sigma_n}\tilde{\mathfrak{h}}_{\lambda_m \sigma_n}\mathfrak{a}_{pp}+\tilde{\mathfrak{h}}_{pp}-2\tilde{\mathfrak{a}}^{\lambda_{m}\sigma_n}\mathfrak{a}_{\lambda_{m}p}\tilde{\mathfrak{h}}_{p \sigma_n})\nonumber\\
+\sqrt{-|\tilde{\mathfrak{a}}|}\frac{\partial}{\partial X^{\rho}}  ( \tilde{\mathfrak{a}}^{\lambda_m \sigma_n}\tilde{\mathfrak{h}}_{\lambda_m \sigma_n}\mathfrak{a}_{pp})+\sqrt{-|\tilde{\mathfrak{a}}|}\frac{\partial}{\partial X^{\rho}}(\tilde{\mathfrak{h}}_{pp}-2\tilde{\mathfrak{a}}^{\lambda_{m}\sigma_n}\mathfrak{a}_{\lambda_{m}p}\tilde{\mathfrak{h}}_{p \sigma_n})=0\label{ee24}
\end{eqnarray}
where, we have used the constraint (\ref{ee23}) coming from tachyon kinematics. 
\subsection{$ D(p-1) $ brane dynamics}
We start with the following action for BPS $ D(p-1) $ brane,
\begin{eqnarray}
\mathcal{S}_{D(p-1)}=-\tau_{p-1}\int d^{p}\xi~\sqrt{-|\tilde{\mathcal{A}}|}+\tau_{p-1}\int \tilde{C}^{(p)}\label{e27}
\end{eqnarray}
where the each of the individual entities in (\ref{e27}) could be formally expressed as,
\begin{eqnarray}
\tilde{\mathcal{A}}_{\alpha_{m}\beta_{n}}& =& \mathcal{G}_{\mu \nu}\partial_{\alpha_{m}}X^{\mu}\partial_{\beta_{n}}X^{\nu}\nonumber\\
\tilde{C}^{(p)}& = & \frac{1}{p!}\tilde{\mathcal{C}}_{\alpha_{0}..\alpha_{p-1}}dX^{\alpha_0}\wedge ...\wedge dX^{\alpha_{p-1}}.
\end{eqnarray}

Using (\ref{E3}), (\ref{E5}) and (\ref{E13}) it is straightforward to compute,
\begin{eqnarray}
\tilde{\mathcal{A}}_{\alpha_{m}\beta_{n}}=\omega^{2}\tilde{\mathfrak{a}}_{\alpha_{m}\beta_{n}}+\tilde{\mathfrak{h}}_{\alpha_{m}\beta_{n}}+\frac{1}{4\omega^{2}}\tilde{\mathfrak{c}}_{\alpha_{m}\beta_{n}}.\label{E29}
\end{eqnarray}

Finally, we note down the RR $ p $ form corresponding to BPS $ D(p-1) $ brane \cite{Bergshoeff:2015uaa},
\begin{eqnarray}
\tilde{\mathcal{C}}_{\alpha_{0}..\alpha_{p-1}}=\left(\omega \tau_{\alpha_{0}}^{a_0}-\frac{1}{2\omega}m_{\alpha_{0}} ^{a_0}\right)...\left(\omega \tau_{\alpha_{p-1}}^{a_{p-1}}-\frac{1}{2\omega}m_{\alpha_{p-1}}^{a_{p-1}}\right) \epsilon_{a_{0}...a_{p-1}}.\label{E30}
\end{eqnarray} 
 
 Substituting (\ref{E29}) and (\ref{E30}) into (\ref{e27}) and taking the NC ($ \omega \rightarrow \infty $) limit it is straightforward to obtain the corresponding finite world-volume action,
 \begin{eqnarray}
 \mathcal{S}_{D(p-1)}=-\frac{\tilde{\tau}_{p-1}}{2}\int d^{p}\xi \sqrt{-|\tilde{\mathfrak{a}}|}\tilde{\mathfrak{a}}^{\alpha_{m}\beta_{n}}\tilde{\mathfrak{h}}_{\alpha_{m}\beta_{n}}\label{E31}
 \end{eqnarray}
where, $ \tilde{\tau}_{p-1}=\tau_{p-1}\omega^{p-2} $ is the re-scaled $ D(p-1) $ brane tension. From (\ref{E31}), one can readily obtain the equation of motion corresponding to embedding modes living on the world-volume of $ D(p-1) $ brane,
\begin{eqnarray}
\frac{1}{2}\sqrt{-|\tilde{\mathfrak{a}}|}\left(\frac{\partial \tilde{\mathfrak{a}}_{\alpha_m \beta_n}}{\partial X^{\rho}} \right) \tilde{\mathfrak{a}}^{\alpha_m \beta_n}
\tilde{\mathfrak{a}}^{\lambda_m \sigma_n}\tilde{\mathfrak{h}}_{\lambda_m \sigma_n}+\sqrt{-|\tilde{\mathfrak{a}}|}\frac{\partial}{\partial X^{\rho}}  ( \tilde{\mathfrak{a}}^{\alpha_m \beta_n}\tilde{\mathfrak{h}}_{\alpha_m \beta_n})=0
\end{eqnarray}
which matches precisely to that with (\ref{ee24}) subjected to the following constraints,
\begin{eqnarray}
\mathfrak{a}_{pp}=1 ;~~\tilde{\mathfrak{h}}_{pp}=2\tilde{\mathfrak{a}}^{\lambda_{m}\sigma_n}\mathfrak{a}_{\lambda_{m}p}\tilde{\mathfrak{h}}_{p \sigma_n}.\label{e31}
\end{eqnarray}
\section{Stress energy tensor}
A further confirmation of our result comes from the computation of stress energy tensor corresponding to non BPS $ Dp $ branes where we show that the tachyon kink (\ref{e18}) has a natural interpretation in terms of a lower dimensional BPS $ D(p-1) $ brane configuration. In order to prove this equivalence we start with the finite world-volume action (\ref{e13}),
\begin{eqnarray}
\mathcal{S}_{NDp}=-\frac{\tilde{\tau_p}}{2}(-1)^{p}\int d^{p+1}\xi~V(T)\partial_{\alpha_{p}}T~ \sqrt{-| \tilde{\mathfrak{a}}|}~\mathfrak{a}^{\lambda_j  \sigma_k}\mathfrak{g}_{\lambda_j \sigma_k}
\end{eqnarray}
which yields the corresponding \textit{longitudinal} stress energy tensor in the NC limit as,
\begin{eqnarray}
T_{N Dp}^{(L)\mu \nu}&=&\frac{1}{\sqrt{-|\tau|}}\frac{\delta \mathcal{S}_{N Dp}}{\delta \tau_{\mu \nu}}\nonumber\\
&=&-\frac{\tilde{\tau}_{p-1}}{2}\int~\frac{ d^{p}\xi}{\sqrt{-| \tau |}}\sqrt{-| \tilde{\mathfrak{a}}|}\left(\frac{1}{2}\partial_{\alpha_{m}}X^{\mu}\partial_{\beta_{n}}X^{\nu}\tilde{\mathfrak{a}}^{\alpha_{m}\beta_n}\tilde{\mathfrak{a}}^{\lambda_{m}\sigma_n}\tilde{\mathfrak{h}}_{\lambda_{m}\sigma_n}+\frac{\delta \tilde{\mathfrak{a}}^{\lambda_m \sigma_n}}{\delta \tau_{\mu \nu}}\tilde{\mathfrak{h}}_{\lambda_m \sigma_n} \right)\nonumber\\
& \equiv &T_{ D(p-1)}^{(L)\mu \nu}.
\end{eqnarray}
subjected to the above constraints (\ref{ee23}) and (\ref{e31}). Here, $ T_{ D(p-1)}^{(L)\mu \nu} $ is the stress tensor corresponding to BPS $ D(p-1) $ brane configuration (\ref{E31}) together with the fact that,
\begin{eqnarray}
\tilde{\tau}_{p-1}&=&(-1)^{p}\int d\xi^{p}~\frac{V(\mathfrak{f})}{\zeta\mathfrak{f}'} 
\end{eqnarray}
is the the effective $ D(p-1) $ brane tension.

Finally, using the above constraints (\ref{ee23}) and (\ref{e31}) and following similar steps it is quite trivial to compute the stress energy tensor corresponding to transverse metric ($ \tilde{\mathfrak{h}}_{\mu \nu}:= \mathfrak{h}_{\mu \nu}+\mathfrak{b}_{\mu \nu}$),  
\begin{eqnarray}
T_{N Dp}^{(T)\mu \nu}&=&\frac{1}{\sqrt{-|\tilde{\mathfrak{h}}|}}\frac{\delta \mathcal{S}_{N Dp}}{\delta \tilde{\mathfrak{h}}_{\mu \nu}}\nonumber\\
&=&-\frac{\tilde{\tau}_{p-1}}{2}\int~\frac{ d^{p}\xi}{\sqrt{-|\tilde{ \mathfrak{h}} |}}\sqrt{-| \tilde{\mathfrak{a}}|}\tilde{\mathfrak{a}}^{\alpha_{m}\beta_{n}}\partial_{\alpha_{m}}X^{\mu}\partial_{\beta_{n}}X^{\nu}\nonumber\\
& \equiv &T_{ D(p-1)}^{(T)\mu \nu}
\end{eqnarray}
where, $ T_{ D(p-1)}^{(T)\mu \nu} $ is the (\textit{transverse}) stress energy tensor corresponding to BPS $ D(p-1) $ brane configuration (\ref{E31}) in the NC background.

\section{Summary and final remarks}
We conclude with a brief summary of the analysis performed. The primary motivation behind our present analysis was to explore the physics of tachyon condensation over string Newton-Cartan (NC) geometry by probing the background with non BPS $ Dp$ brane. Unlike the case for BPS $ Dp $ branes, we notice that the effective WZ coupling corresponding to non BPS $ Dp $ branes has to scale linearly with the NC parameter ($ \omega $) in order for the world-volume theory to make sense in the NC ($ \omega \rightarrow \infty $) limit. This stems from the fact that non BPS $ Dp $ branes couple with background RR $ p $ form rather than a $ (p+1) $ form.

Considering the most generic embedding for the world-volume scalars we show that the spatial dependent tachyon condensation leads to an emerging BPS $ D(p-1) $ brane dynamics over string NC geometry. We further solidify our claim by computing stress energy tensor(s) associated with non BPS $ Dp $ brane and show that the resulting dynamics could be thought of as being that of the reminiscent of BPS $ D(p-1) $ brane configuration over string NC background.\\ 

{\bf {Acknowledgements :}}
The author is indebted to the authorities of IIT Roorkee for their unconditional support towards researches in
basic sciences. \\ 

\end{document}